\documentclass[a4paper,11pt]{article}

\usepackage{amsmath}
\usepackage{amssymb}
\usepackage{amsthm}
\usepackage{array}
\usepackage{indentfirst}
\usepackage{graphicx}
\usepackage[utf8]{inputenc}
\usepackage[dvipsnames]{xcolor}
\usepackage{tabularx}
\usepackage{caption}
\usepackage{subcaption}
\usepackage{gensymb}
\usepackage[normalem]{ulem}
\usepackage{array}
\usepackage{enumitem}

\usepackage{amssymb}

\usepackage{jcappub}
\newcolumntype{Y}{>{\centering\arraybackslash}X}
\newcolumntype{W}[1]{>{\centering\arraybackslash\hsize=#1\hsize}X}

\hypersetup{colorlinks=true, linkcolor=blue, urlcolor=blue, citecolor=blue, linktocpage=true}

\begin{document}

\title{Searching for EeV photons with Telescope Array Surface Detector and neural networks.}

\author[1]{R.U. Abbasi,}
\author[1,2]{T. Abu-Zayyad,}
\author[2]{M. Allen,}
\author[2]{J.W. Belz,}
\author[2]{D.R. Bergman,}
\author[3]{F. Bradfield,}
\author[2]{I.  Buckland,}
\author[2]{W. Campbell,}
\author[4]{B.G. Cheon,}
\author[3]{K. Endo,}
% ...
\collaboration{Telescope Array Collaboration}

\affiliation[1]{Department of Physics, Loyola University-Chicago, Chicago, Illinois 60660, USA}
\affiliation[2]{High Energy Astrophysics Institute and Department of Physics and Astronomy, University of Utah, Salt Lake City, Utah 84112-0830, USA}
\affiliation[3]{Graduate School of Science, Osaka Metropolitan University, Sugimoto, Sumiyoshi, Osaka 558-8585, Japan}
\affiliation[4]{Department of Physics and The Research Institute of Natural Science, Hanyang University, Seongdong-gu, Seoul 426-791, Korea}
\affiliation[5]{Institute of Physics, Academia Sinica, Taipei City 115201, Taiwan}
\affiliation[6]{Institute for Cosmic Ray Research, University of Tokyo, Kashiwa, Chiba 277-8582, Japan}
\affiliation[7]{Nambu Yoichiro Institute of Theoretical and Experimental Physics, Osaka Metropolitan University, Sugimoto, Sumiyoshi, Osaka 558-8585, Japan}
\affiliation[8]{CEICO, Institute of Physics, Czech Academy of Sciences, Prague 182 21, Czech Republic}
\affiliation[9]{Institute of Astronomy, National Autonomous University of Mexico Ensenada Campus, Ensenada, BC 22860, Mexico}
\affiliation[10]{Graduate School of Engineering, Osaka Electro-Communication University, Neyagawa-shi, Osaka 572-8530, Japan}
\affiliation[11]{Faculty of Engineering, Kanagawa University, Yokohama, Kanagawa 221-8686, Japan}
\affiliation[12]{Astrophysical Big Bang Laboratory, RIKEN, Wako, Saitama 351-0198, Japan}
\affiliation[13]{Department of Physics, Sungkyunkwan University, Jang-an-gu, Suwon 16419, Korea}
\affiliation[14]{Department of Physics, Tokyo City University, Setagaya-ku, Tokyo 158-8557, Japan}
\affiliation[15]{Institute for Nuclear Research of the Russian Academy of Sciences, Moscow 117312, Russia}
\affiliation[16]{Faculty of Systems Engineering and Science, Shibaura Institute of Technology, Minumaku, Tokyo 337-8570, Japan}
\affiliation[17]{Academic Assembly School of Science and Technology Institute of Engineering, Shinshu University, Nagano, Nagano 380-8554, Japan}
\affiliation[18]{Service de Physique Théorique, Université Libre de Bruxelles, Brussels 1050, Belgium}
\affiliation[19]{Department of Physics, Yonsei University, Seodaemun-gu, Seoul 120-749, Korea}
\affiliation[20]{Center for Astrophysics and Cosmology, University of Nova Gorica, Nova Gorica 5297, Slovenia}
\affiliation[21]{Faculty of Science, Kochi University, Kochi, Kochi 780-8520, Japan}
\affiliation[22]{College of Science and Engineering, Chubu University, Kasugai, Aichi 487-8501, Japan}
\affiliation[23]{Quantum ICT Advanced Development Center, National Institute for Information and Communications Technology, Koganei, Tokyo 184-8795, Japan}
\affiliation[24]{Doctoral School of Exact and Natural Sciences, University of Lodz, Lodz, Lodz 90-237, Poland}
\affiliation[25]{Sternberg Astronomical Institute, Moscow M.V. Lomonosov State University, Moscow 119991, Russia}
\affiliation[26]{Department of Physics, School of Natural Sciences, Ulsan National Institute of Science and Technology, UNIST-gil, Ulsan 689-798, Korea}
\affiliation[27]{Astrophysics Division, National Centre for Nuclear Research, Warsaw 02-093, Poland}
\affiliation[28]{Earthquake Research Institute, University of Tokyo, Bunkyo-ku, Tokyo 277-8582, Japan}
\affiliation[29]{Graduate School of Information Sciences, Hiroshima City University, Hiroshima, Hiroshima 731-3194, Japan}
\affiliation[30]{Institute of Particle and Nuclear Studies, KEK, Tsukuba, Ibaraki 305-0801, Japan}
\affiliation[31]{School of Science and Engineering, Tokyo Denki University, Saitama 350-0394, Japan}

\emailAdd{ivan.kharuk@phystech.edu}

\abstract{
Ultra-high-energy photons play an important role in probing astrophysical models and beyond-Standard-Model scenarios.
We report updated limits on the diffuse photon flux using Telescope Array's Surface Detector data collected over 14 years of operation. 
Our method employs a neural network classifier to effectively distinguish between proton-induced and photon-induced events.
The input data include both reconstructed composition-sensitive parameters and raw time-resolved signals registered by the Surface Detector stations.
To mitigate biases from Monte Carlo simulations, we fine-tune the network with a subset of experimental data.
The number of observed photon candidates is found to be consistent with the expected hadronic background, yielding upper limits on photon flux $\Phi_\gamma(E_\gamma > 10^{19} \text{eV}) < 2.3 \cdot 10^{-3} $, and $\Phi_\gamma(E_\gamma > 10^{20} \text{eV}) < 3.0 \cdot 10^{-4} $ $ (\text{km}^2 \cdot \text{sr} \cdot \text{yr})^{-1} $.
}

\maketitle

\section{Introduction}
\label{sec:introduction}

Cosmogenic ultra-high-energy (UHE) photons, with energies above $ 10^{18} $ eV, are of interest for two main reasons. First, they can be produced via the Greisen-Zatsepin-Kuzmin (GZK) mechanism~\cite{greisen1966end,Zatsepin:1966jv, Wdowczyk:1971jk,Michalak:1990iw,Gelmini:2007jy, Hooper:2010ze, Gelmini:2022evy} in the interstellar medium. The resulting photon flux depends on the mass composition of cosmic rays, thus providing an independent tool for their analyses. Second, registering ultra-high-energy photons beyond the expected GZK flux level would provide strong evidence for models beyond the Standard Model of particle physics \cite{berezinsky1998signatures, Anchordoqui:2021crl, Fairbairn:2009zi}, including, in particular, decaying dark matter \cite{Berezinsky:1997hy, Kuzmin:1997jua, Kuzmin:1999zk,Kalashev:2016cre,Kalashev:2020hqc,Chianese:2021jke, Das:2023wtk} and violation of Lorentz symmetry \cite{coleman1999high, galaverni2008lorentz, rubtsov2014prospective,PierreAuger:2021tog}.

Experimental searches for ultra-high-energy photons rely on complementary detection techniques, each with distinct advantages and limitations. Fluorescence detectors measure fluorescence light from extensive air showers (EAS), allowing for the direct observation of the longitudinal shower development and the depth of shower maximum ($X_{max}$). This provides excellent discrimination between photon and hadron primaries, but their operation time is restricted to clear, moonless nights, resulting in a limited duty cycle that constrains the exposure accumulation for ultra-high-energy events. Conversely, scintillator-based surface detectors, measuring secondary particles at the ground level, operate continuously, offering the vast exposure required to probe low-flux regimes. However, they face challenges in distinguishing muon and electromagnetic components of EAS, whose ratio is the key for distinguishing hadron-induced and photon-induced EAS. To address this, experiments like Yakutsk~\cite{Glushkov:2009tn} employed dedicated muon detectors to suppress the hadronic background. Similarly, the Pierre Auger Observatory has set stringent limits in the Southern Hemisphere by utilizing both hybrid observations~\cite{PierreAuger:2024ayl} and water-Cherenkov Surface Detectors~\cite{abreu2023search}, which possess different sensitivities to the muon component compared to scintillators. Early constraints were also established by the AGASA experiment~\cite{Shinozaki:2002ve}. 

A different approach was employed in the Telescope Array (TA) experiment. TA is a large-scale cosmic-ray experiment located in Utah, USA, 39\degree 17' 49'' N 112\degree54' 31'' W, at approximately 1 400 m above sea level \cite{abu2012surface, tokuno2012new, udo2017telescope}. It aims to study ultra-high-energy cosmic rays (UHECR) using its Surface Detector (SD) and Fluorescence Detector (FD). Since high exposure is crucial for probing low fluxes, TA used its SD data to set an upper limit on the photon flux using composition-sensitive parameters, such as EAS front curvature \cite{TelescopeArray:2013yze}, which were further improved using boosted decision trees \cite{TelescopeArray:2018rbt}.

To date, no experiments have reported a significant excess of photon-induced events above the expected hadronic background (see Figure \ref{img_results}). This background arises from the imperfect discrimination between photon and hadron primaries, driven by the inherent stochasticity of air-shower evolution. Specifically, hadron-induced showers can fluctuate to develop a dominant electromagnetic component. A primary example is the production of a leading $\pi^0$ that decays via $\pi^0 \rightarrow \gamma\gamma$ channel \cite{Pawlowsky:2024pth}; such events are virtually indistinguishable from primary photon showers and constitute an irreducible background.

In this paper, we search for ultra-high-energy photons in 14 years of TA SD data using a neural network classifier, capable of leveraging complex patterns in the data to discern between photon-induced and proton-induced events. In the field of Astroparticle Physics, neural networks have been successfully applied in TA \cite{Kalashev:2020fzg,Ivanov:2020nfo,Kalashev:2021vop}, the Pierre Auger Observatory \cite{PierreAuger:2021fkf, PierreAuger:2023gmj, Glombitza:2020yhw}, KASCADE \cite{Kuznetsov:2023kss, Kuznetsov:2023pvo}, and neutrino telescopes \cite{Huennefeld:2017pdh, IceCube:2021umt, DeSio:2019lcr, Spisso:2022tpm, Kharuk:2022abi, Kharuk:2023xnl}, illustrating the potential of machine learning for our task. 

The present analysis is the natural evolution of \cite{Kharuk:2023gts}, which demonstrated the viability of a neural-network-based photon search for TA SD data. The novelty of our approach is that we ensure that predictions of our neural network on the experimental data are reliable by mitigating the discrepancies between Monte Carlo simulations and the experimental data. For this purpose our technique includes a procedure for fine-tuning the neural networks using a relevant subset of the experimental data. Our method also incorporates a blind optimization of the neural network classification threshold to establish the most stringent constraint on the photon flux. 

\begin{figure}
\centering
  \begin{minipage}[b]{0.475\textwidth}
  \center{\includegraphics[width=1.\linewidth]{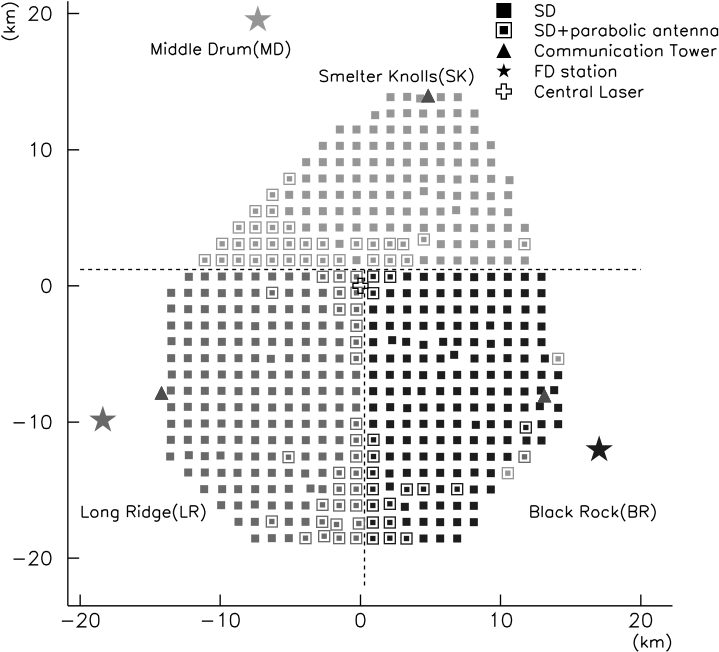}}
  \caption{Schematic representation of the arrangement of the TA SD and FD stations, \cite{abu2012surface}.}
  \label{scheme_view_ta}
  \end{minipage}
\hfill
  \begin{minipage}[b]{0.475\textwidth}
  \center{\includegraphics[width=1.\linewidth]{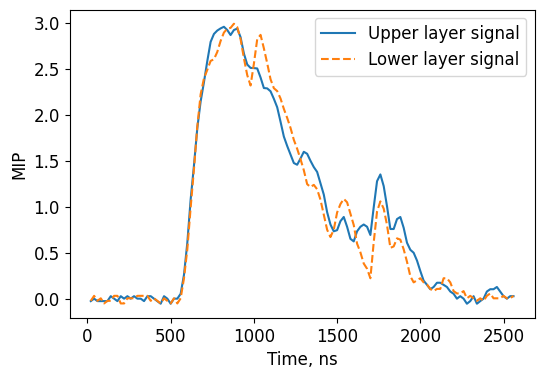}}
  \caption{An example of processed and calibrated waveforms registered by a TA SD station. Blue and orange lines depict signals from the upper and lower scintillator layers, respectively.}
  \label{wf_example}
  \end{minipage}
\end{figure}

The neural network operates directly on the signals recorded by the TA SD stations, whose main features we now summarize. The TA SD array, depicted in Figure \ref{scheme_view_ta}, comprises 507 above-ground scintillation stations with 1.2 km spacing, arranged in a square grid covering an area of approximately 700 km$^2$. Each SD station consists of two layers of plastic scintillator, which register signals with a time resolution of 20 ns \cite{abu2012surface}. All stations are calibrated in real time using atmospheric muons, with the signal measured in conventional units called a ``minimal ionizing particles'' (MIPs). For our analysis, we retain 128 consecutive time bins from each layer of triggered SD stations, which we refer to as waveforms. An example of a waveform registered by a single SD station is presented in Figure \ref{wf_example}.

The sensitivity threshold of the TA SD for photon-induced events is approximately $10^{18.1}$~eV (see Figure~\ref{en_spectrum}). While this allows us to derive an integral limit for $E > 10^{18}$~eV, the very limited exposure between $10^{18}$ and $10^{18.5}$~eV implies that such a limit would be dominated almost entirely by higher-energy photons, and hence could be misleading. We therefore present results only for $E > 10^{18.5}$~eV, where the TA SD has sufficient direct sensitivity.

\section{Method}
\label{sec:method}

We start by outlining the general scheme of the analysis and stating our main assumptions. The method comprises four main steps, which we describe below.

1. \textit{Data selection and preparation.} We perform Monte Carlo (MC) simulations of 14 years of TA SD data, modeling air showers initiated by protons and photons (see Section \ref{sec:mc} for details). Heavy nuclei such as iron produce showers that develop earlier, exhibit denser cores, and higher muon-to-electromagnetic ratios than proton-induced cascades \cite{Ros:2011zg, ros2013improving, cazon2023universality}. This makes them easier to distinguish from photon-induced air showers, yielding protons as the main background source to photons signal.

2. \textit{Neural network training.} We train a neural network classifier to distinguish between proton- and photon-induced EAS events using a MC-simulated data set. For each event the neural network outputs a number $ \xi \in [0;1] $ representing its confidence that a given event is proton-like (closer to 0) or photon-like (closer to 1).

Monte Carlo simulations might have slight differences from experimental data that are irrelevant for standard algorithmic reconstruction but are utilized by the neural network. To account for this fact, we fine-tune the neural network using a subset of the experimental data confidently identified as proton-induced events (see Section \ref{sec:training} for details). This step allows the neural network to learn from actual data, thus reducing MC-specific biases. The data used at this step is not used in the subsequent computation of the photon flux upper limit. In what follows, we refer to this set as the burn sample.

3. \textit{Blind optimization of the classification threshold.} At present, the most energetic photons observed in the experiment have energies of the order of PeV~\cite{TibetASgamma:2021tpz, LHAASO:2021gok}. Strong constraints have been set on the upper limit of photon flux at EeV energies~\cite{Fomin:2017ypo, KASCADEGrande:2017vwf, PierreAuger:2025jwt, aab2017search, abreu2023search, TelescopeArray:2018rbt} without detecting photon signal --- the number of observed photon candidates was consistent with background expectations. Based on this, we optimize the neural network's classification threshold to provide the strongest constraint on the photon flux before unblinding the data and calculating the actual number of photon candidates.

For a given classification threshold $\xi$, the 95\% upper limit on photon flux $\Phi$ can be calculated as the ratio of 95\% confidence level on the number of observed photon candidates, $\sigma^{95}(n^{\text{cand}}(\xi))$, to the effective TA SD exposure, $\Omega(\xi)$:
\begin{equation}
\label{upper-limit}
\Phi_{\gamma} = \frac{\sigma^{95}(n^{\text{cand}}(\xi))}{\Omega(\xi)} \;.
\end{equation}
We minimize this value by performing blind optimization of the classification threshold on the MC data sets. For protons, we assume HiRes \cite{HiRes:2007lra} energy spectrum, which is close to experimental estimates. For photon primaries we choose $E^{-2}$ differential energy spectrum, following common choice in similar studies \cite{TelescopeArray:2023bdy, PierreAuger:2024ayl, abreu2023search}.

The effective TA SD exposure is the product of the following factors: data collection time; product of area and solid angle used in the MC simulation; trigger efficiency (ratio of events that invoked TA SD trigger to the total number of MC events); event selection efficiency (fraction of events passing quality cuts after the reconstruction); and photons selection efficiency, $ S(\xi) = \frac{n_\gamma (\xi)}{ n_\gamma (0) }$, where $ n_\gamma (\xi) $ is the number of photons passing neural network's classification threshold. Note that only photon selection efficiency depends on $\xi$. Hence all other factors can be omitted for the classification threshold optimization.

The estimation of the numerator in Equation \ref{upper-limit} requires a multi-step procedure. Initially, the proton suppression level, $P(\xi)$, is quantified on the MC sample. It is defined as the ratio of events incorrectly identified as photons to the total number of proton events for a given $\xi$. Subsequently, the expected number of false photons due to the proton background in the experimental data, $n^{\text{bg}}(\xi)$, is estimated by multiplying $P(\xi)$ by the total number of experimental events. Further, the corresponding $\mathrm{95\%}$ confidence level upper limit is determined using the Feldman-Cousins statistics \cite{Feldman:1997qc} under the assumption of zero background, which is designated as $\sigma^{95}(n^{\text{bg}}(\xi))$. Given that event registration is inherently a random process, $\mathbf{n^{\text{bg}}(\xi)}$ should be considered as a Poisson random variable, hence optimizing $\sigma^{95}(n^{\text{bg}}(\xi))$ expectation value.

Putting all together, we fix $\xi_{\text{opt}}$ on MC proton and photon data sets by minimizing
\begin{equation} \label{merit_func}
    L^{95} = \frac{ \textit{M} \left( \sigma^{95}(n^{\text{bg}}(\xi)) \right) }{ S(\xi) }  ~ \,, 
\end{equation}
where $ M(\cdot) $ stands for the expectation value.

4. \textit{Obtaining photon flux limits.} As the final step of the analysis, we unblind the data and count the number of photons candidates that passed the classification threshold optimized in the previous step. This gives $n^{\text{cand}}(\xi_{\text{opt}})$, substituting which into Eq. \ref{upper-limit} yields the resulting limit on the diffuse photon flux.

In the following sections, we describe in detail the data and Monte Carlo sets along with the neural network architecture and training. The final results are presented in Section \ref{sec:results}.

\section{Data sets and Monte Carlo simulations}
\label{sec:mc}

In this study we use TA SD data recorded during 14 years of observation between 2008-05-11 and 2022-05-10. We process the data and MC simulations using the same reconstruction procedure and apply the same event quality cuts, as described below.

MC simulations conducted for this study encompass the complete evolution of air showers and the subsequent response of TA SD stations. These simulations are performed to model 14 years of TA operation, using CORSIKA version 7.7420 \cite{Heck:1998vt} for air-shower simulations and GEANT4 \cite{GEANT4:2002zbu} for detector response modeling. The simulations randomly sample event timestamps and reproduce the corresponding experimental conditions by using actual real-time calibration tables and list of non-functional TA SD stations, thus ensuring an accurate representation of the detector behavior.

For proton-induced events, we employed three high-energy hadronic interaction models: QGSJET-II-04~\cite{ostapchenko2006qgsjet}, EPOS-LHC~\cite{pierog2009epos}, and Sibyll 2.3d~\cite{ahn2009cosmic}. Low-energy hadronic and electromagnetic interactions were simulated using FLUKA~\cite{ferrari2005fluka} and EGS4 \cite{nelson1985egs4}, correspondingly. 

The simulations were performed in three energy ranges, starting from $ 10^{17.45},~10^{18.95}, $ and $ 10^{19.45} $~eV. Within each sample, the proton energy spectrum is modeled to follow the one measured by the HiRes experiment \cite{HiRes:2007lra}. Due to the sharp drop-off of the spectrum at high energies, such splitting is required for a comprehensive comparison of the MC and experimental data at different energy scales. These datasets were subsequently used for setting upper limits on photon flux at different energy cutoffs. The number of simulated air showers is the same for all of the high-energy hadronic interaction models.

Photon-induced events are simulated using the QGSJET-II-04, FLUKA, and EGS4. We use the PRESHOWER code \cite{Homola:2003ru} to account for geomagnetic interactions resulting in magnetic $e^+$,~$e^-$ pair production. The corresponding air showers are dominated by their electromagnetic component, hence the effect of choosing a particular high-energy hadron interaction model should be negligible. We test this assumption with Monte Carlo simulations in Section~\ref{sec:model_dep}. The differential energy spectrum is chosen to be $E^{-1}$ to cover the whole range of energies within a single data set, thus avoiding underrepresentation of high-energy events.

MC and experimental data were passed through a reconstruction procedure \cite{TelescopeArray:2013yze} optimized to infer the parameters of photon-induced events. Since we are searching for photon-like events, we estimate the energy of events based on the look-up table constructed using MC photon data set \cite{TelescopeArray:2018rbt}, which is a function of the reconstructed $S_{800}$ value (scintillator signal density at the distance of 800 m from the core), zenith and azimuth angles. Thus obtained energy, $E_\gamma$, corresponds to average energy of the primary photon, inducing a shower with the same arrival direction and $S_{800}$. In contrast to proton-optimized energy reconstruction, this approach ensures that we set corrects limits on the photon flux for a specified energy threshold. For proton primaries, $E_\gamma$ yields systematically higher energy estimate than $E_p$, an analogous unbiased energy estimator for the proton primaries. For instance, at $\sim 10$~EeV Monte-Carlo energy the mean $E_\gamma$ for proton primaries is $\sim 55\%$ higher than their $E_p$. The full list of reconstructed features, presented in Appendix \ref{app:reco}, includes geometrical and primary particle mass-sensitive parameters, such as reconstructed zenith angle and Linsley front curvature.

To ensure good quality of events reconstruction, we applied the following selection criteria to both the MC and the experimental data sets:
\begin{itemize}[noitemsep,topsep=0pt]
    \item Reconstructed zenith angle is below 55\degree;
    \item Minimum of 7 triggered stations per event;
    \item Reconstructed shower core position is within the SD array, at least 1200 meters from its boundary;
    \item Joint geometry and lateral distribution profile fit accuracy of $ \chi^2 / d.o.f. < 5 $.
\end{itemize}

The chosen set of event quality cuts is close to the standard ones used for anisotropy studies with proton primaries in TA, providing a balance between reconstruction accuracy and photon exposure. The cut on the reconstructed shower core position ensures that geometry and charge deposition profiles are well-understood, without extrapolating peak values outside TA SD array. Note that the cut in the zenith angle is slightly stricter than in our previous studies~\cite{TelescopeArray:2018rbt, TelescopeArray:2019mwd}. Additionally, since it is impossible to properly simulate the signal in the saturated stations, we exclude them both from MC-simulated and experimental data.

\begin{figure}
\center{\includegraphics[width=0.5\linewidth]{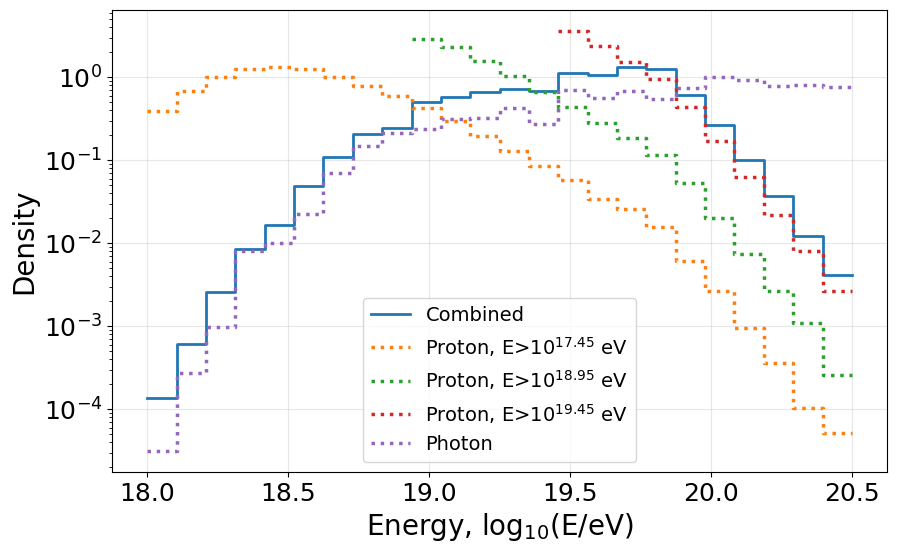}}\\
\caption{Energy distribution of the MC samples after applying event selection criteria. Dotted lines show energy distributions of each individual data set. The \textit{combined} sample, which as equal mixture of proton-induced and photon-induced events with the same energy spectrum, was used for neural network training.}
\label{en_spectrum}
\end{figure}

To create a balanced data set for neural network training and evaluation, we employ a mixing procedure for proton-induced and photon-induced events. The energy range is divided into bins of constant width ($ 0.05 $ in $ \text{log}_{10} $ scale), in each of which the events are randomly sampled so that the total number of registered and successfully reconstructed events of both types is equal. The number of events in each bin is limited by data set with lower event count, up to sampling fluctuations. This makes the energy spectrum of proton- and photon-induced events the same, which mitigates potential biases related to event energy, which could otherwise be exploited by the neural network as a discriminative yet unphysical parameter. The resulting energy spectrum of events passing event quality cuts in MC sample is presented in Figure~\ref{en_spectrum}.

The final MC sample comprises approximately $ 5 \cdot 10^6 $ events in total, which is mixture of proton primaries for all three hadronic interaction models and photon primaries. Events were partitioned into training, test, and validation sets in a ratio of 8:1:1. The test data set is utilized for optimizing neural network parameters, while the validation data set serves for blind optimization of the classification threshold and estimation of TA SD exposure.

For the experimental data, we have excluded all lightning-correlated events. That is, we reject events registered within a 10-minute interval related to lightnings recorded by the National Lightning Detection Network \cite{cummins2009overview} at the location of the TA SD. This step is necessary as lightning can be registered by TA SD as photon-like events \cite{abbasi2017bursts}.

\section{Neural network}
\label{sec:nn}

\subsection{Architecture}
\label{sec:nn_arch}

The neural network architecture used in this study is a slight modification of the one used previously for the analysis of mass composition \cite{kalashev2022deep} and the estimation of the flux of photons \cite{Kharuk:2023gts}. Its architecture, presented in Figure \ref{nn_arch}, consists of three main components, each focusing on different aspects of the cosmic-ray-induced air shower:

\begin{figure}
\center{\includegraphics[width=0.98\linewidth]{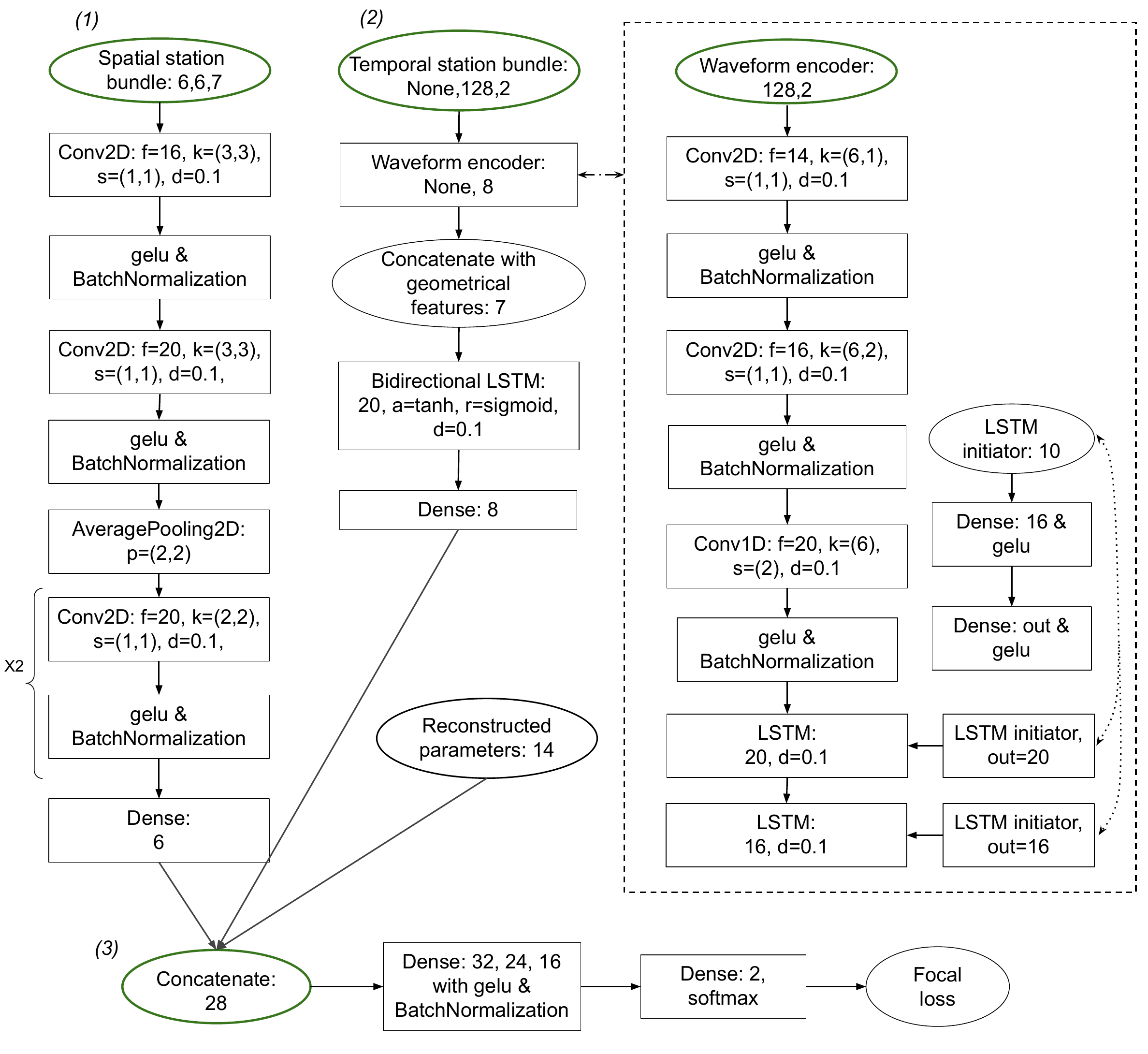}}\\
\caption{Architecture of the neural network used to distinguish proton-induced and photon-induced events: (1) -- spatial station bundle, (2) -- temporal station bundle, and (3) -- combined analysis. The waveform encoder architecture, which is a sub-network in the temporal station bundle, is shown inside the dotted box. Solid arrows depict data flow, dotted lines connect elements with their detailed description. Layer names and parameters follow TensorFlow API.}
\label{nn_arch}
\end{figure} 

(1) \textit{Spatial station bundle.} This component analyses the geometric pattern of activated SD stations. It treats each event as an ``image'' on a 6x6 grid of stations centered around the reconstructed shower core. This representation allows the network to identify the spatial features of an air shower. Convolutional neural networks are known to be well suited for analyzing such data, allowing for efficient parsing of the geometrical structure of an event.

For each of the stations on the grid, the input data includes:
\begin{itemize}[noitemsep,topsep=0pt]
    \item Stations coordinates relative to the reconstructed position of the shower core;
    \item Integral charge registered by the station;
    \item Reconstructed time of the shower plane front arrival;
    \item Time difference between the actual station activation time and the plane front arrival;
    \item Mask whether the station was triggered in an event or not. 
\end{itemize}

The splitting of station activation time into two components, items 3 and 4 in the list above, is useful as it provides the neural network the information on the front curvature. The data for the stations that were not triggered in an event are filled with zeros.

\begin{figure}
\center{\includegraphics[width=1.\linewidth]{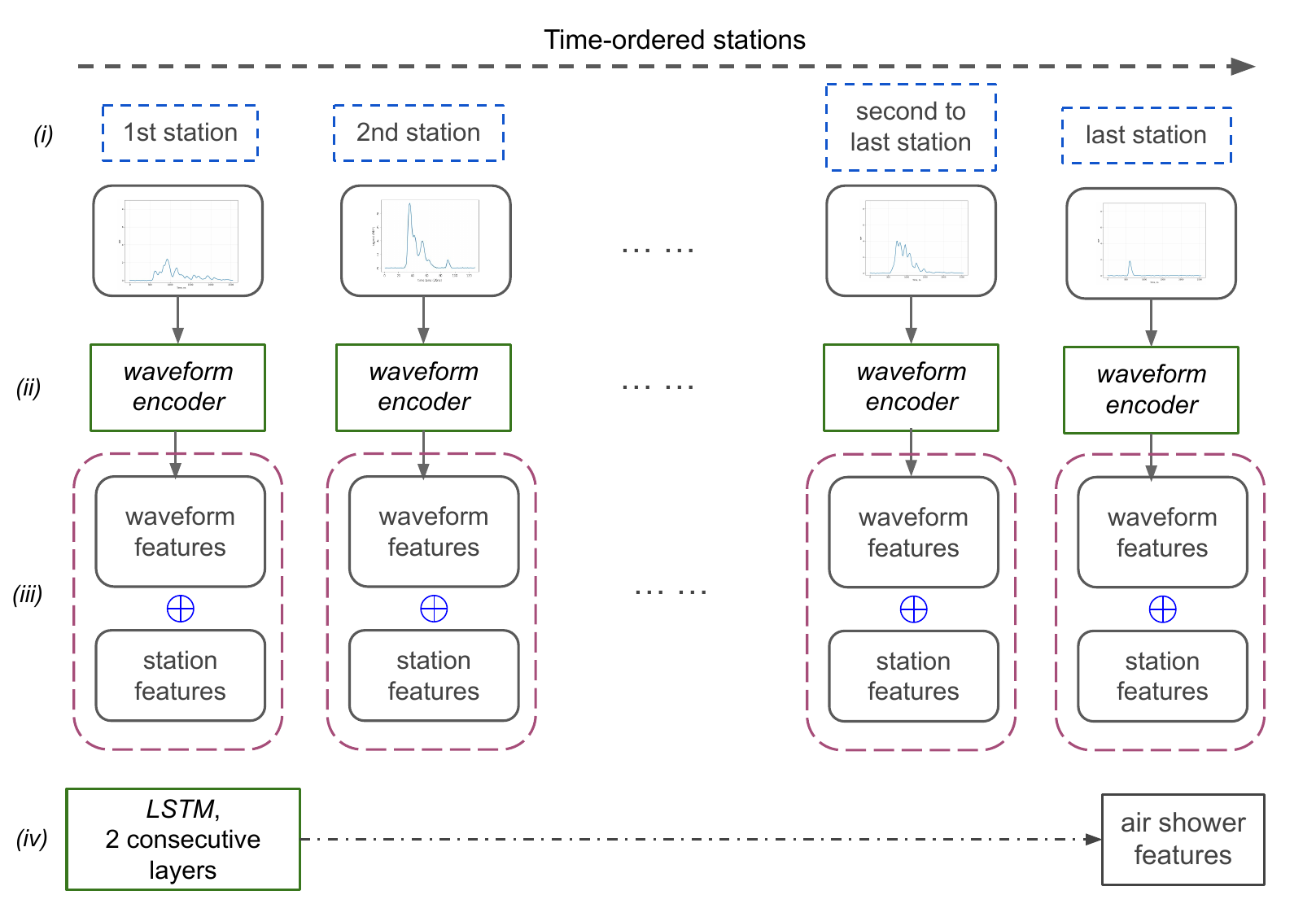}}\\
\caption{Architecture of the temporal station bundle. See the temporal station bundle in Figure \ref{nn_arch} for layers characteristics. For simplicity, we show only one waveform per SD station.}
\label{tmp_bndl_arch}
\end{figure}

(2) \textit{Temporal station bundle.} The triggered stations naturally form a time series when ordered according to their activation times. This sequence captures the structure of an air shower, specifically how it evolves from the periphery to the center and again to the periphery, thus providing the neural network with useful information. The diagram of data processing in this part of the neural network is depicted in Figure \ref{tmp_bndl_arch} and goes as follows:
\begin{itemize}[noitemsep,topsep=0pt]
    \item[(i)] The input data comprises a set of SD stations ordered according to their activation times (red dashed boxes in Figure \ref{tmp_bndl_arch}). Unlike the spatial station bundle, in this part of the neural network we also use raw waveforms registered by the stations. 
    \item[(ii)] Waveforms from upper and lower layers of SD station form a two-channel one-dimensional ``image'', which can also be considered as a sequence. We apply a \textit{waveform encoder} to each of the registered waveforms to extract features representing waveform characteristics using a combination of convolutional and recurrent layers (specifically, long short-term memory (LSTM) blocks). We use \textit{LSTM initiator} sub-network to initialize the state of the recurrent layers, utilizing reconstructed event properties, station coordinates and activation times (see Fig. \ref{nn_arch}).
    \item[(iii)] Second, thus extracted features are concatenated with individual station features -- their coordinates, activation times, and integral registered charges. This yields full set of station features, represented as magenta dashed boxes in Figure \ref{tmp_bndl_arch}.
    \item[(iv)] Finally, the resulting ``station sequence'' is passed through a bidirectional Long Short-Term Memory layer, which outputs event characteristics as a whole.
\end{itemize}

(3) \textit{Combined analysis.} To infer whether an event was proton- or photon-induced, we concatenate the features extracted by the \textit{spatial station bundle} and the \textit{temporal station bundle}. We supplement this data with reconstructed event characteristics, such as primary zenith and azimuth angles, $ S_{800} $, and others (see Appendix \ref{app:reco}). Together, they are passed through a series of fully connected (dense) layers. The output of the last layer is a number $ \xi \in [0;1] $ representing the neural network's confidence that the corresponding event is photon-like. 

In total, the neural network has 29000 trainable parameters. We used the Adam optimizer \cite{kingma2014adam} and employ early stopping for terminating neural network training. The neural network is implemented using the TensorFlow library \cite{tensorflow2015-whitepaper}.

We explored the importance of each of the neural network components described above by training a neural network with only one of them. The temporal station bundle, reconstruction parameters, and spatial station bundle yielded 96\%, 85\%, and 81\% separation accuracy, respectively. We would like to note that it was essential to combine all of these blocks in a single architecture to achieve the best metrics.

\subsection{Training Process}
\label{sec:training}

To optimize the neural network capability for separating proton- and photon-induced events, we employed several training techniques:

(1) Weighted training. We assigned higher weights to proton-induced events during training (specifically, five for protons and one for photons). The choice of class weights is a hyperparameter optimized to enhance the final sensitivity of the analysis. Since our goal is to set the strongest constraints on the photon flux, the optimal estimator should minimize the false-positive rate (proton background misidentified as photons) in the critical signal region ($\xi \approx 1$), even at the cost of overall classification balance. This strategic prioritization of proton rejection is a deliberate feature of the training optimization. It does not bias the final flux estimation, as the validity of the classifier's performance is ensured by the agreement between its predictions on MC-simulated and experimental data.

(2) We used a special loss function, namely, \textit{focal loss} \cite{lin2017focal}:
\begin{equation}
    L = (1-p_{\text{c}})^\gamma \ln{p_{\text{c}}} \;,
\end{equation}
where $ p_{\text{c}} $ is the neural networks confidence for the correct class and $ \gamma $ is a constant, which we fixed to be two. The term $ (1-p_{\text{c}})^\gamma $ modulates the standard cross-entropy loss so that events that were assigned to the correct class with high confidence yield small contribution to the total loss value. Hence neural network pays more attention to events that are difficult to classify.

(3) Ensemble approach. We trained multiple neural networks and selected the top three based on their metrics for constraining the photon flux, Eq. \ref{merit_func}. This approach helps mitigate biases that may arise from the sensitivity of individual networks to specific event features.

Monte Carlo simulations inevitably involve simplifications, such as the application of thinning and de-thinning procedures \cite{Stokes:2011wbu}, as well as uncertainties related to the modeling of air-shower evolution and the response of detector stations. To account for such imperfections, and avoid overfitting to MC data as opposed to experimental data, we employed two complementary strategies.  

First, for neural network training we introduced physically motivated noise into the simulated data, mimicking the resolution of the input parameters. This includes both additive Gaussian noise (40 \textit{ns} for station activation times and 0.1 MIP for waveform signals) and multiplicative noise for measured charge (10\%).

Second, a fine-tuning procedure based on a burn sample of experimental data was implemented. After the initial training of the neural network on MC simulations, the model was used to evaluate the photon-likeness of experimental events. Events that were classified with high confidence ($\xi \leq 0.2$) as proton-induced were included in the training set with corresponding labels. The network was then fine-tuned on this augmented data set, where each training batch contained equal fractions of simulated and experimental proton-induced events. Training was terminated after the burn sample was passed twice through the network, a choice made to prevent overfitting to the limited experimental subset.

Using a burn sample to fine-tune the neural network constitutes the novelty of our approach. It helps the network adapt to the nuances of real data that may not be fully captured in simulations. Importantly, we observed that the effect of using burn sample on neural network metrics is negligible, indicating that there are no major MC-experimental data discrepancies. This step was inspired by surrogate supervision and labeling \cite{Tajbakhsh:8759553}, where a limited subset of data with approximate or indirectly obtained annotations is employed to guide model adaptation to new domains.

\section{Model dependence}
\label{sec:model_dep}

QGSJET-II-04, EPOS-LHC, and Sibyll differently approximate high-energy hadron interactions, which, in particular, results in different muons yield. Since higher muon yield would make easier to distinguish proton-induced and photon-induced events, we investigated the dependence of neural network predictions on the choice of high-energy hadronic interaction model. To this end, we trained two neural networks: one using events simulated with QGSJET-II-04 and EPOS-LHC as high-energy hadronic interaction models and the other using all available models (including Sibyll). For this study, we did not perform fine-tuning of neural networks on experimental data.

\begin{figure}
     \centering
     \begin{subfigure}[b]{0.31\textwidth}
         \centering
         \includegraphics[width=\textwidth]{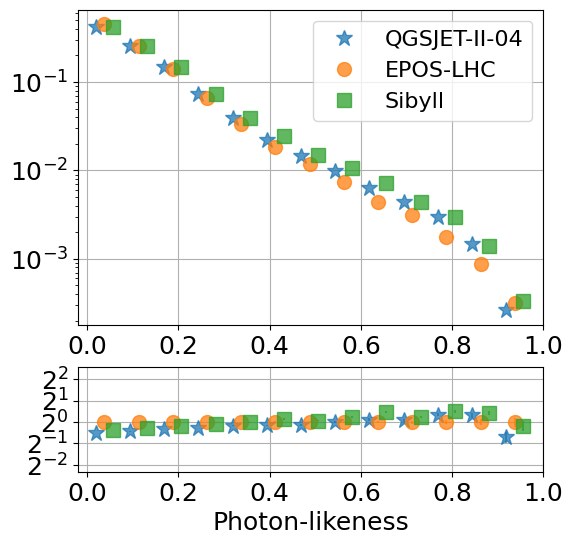}
         \caption{MC $ 10^{17.45} $ eV samples}
         \label{fig:hist_sb_1745}
     \end{subfigure}
     \hfill
     \begin{subfigure}[b]{0.31\textwidth}
         \centering
         \includegraphics[width=\textwidth]{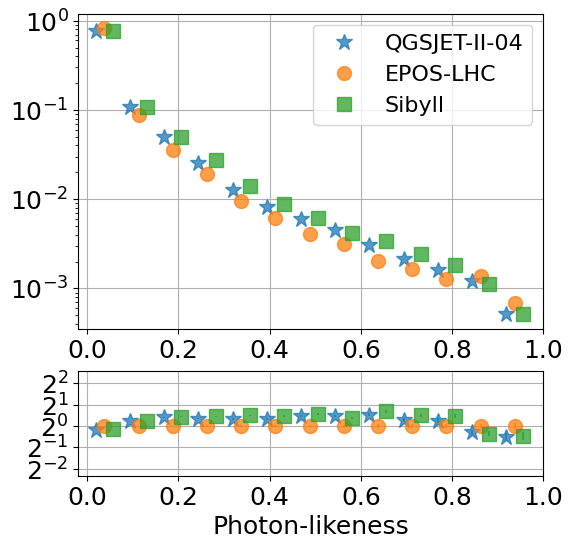}
         \caption{MC $ 10^{18.95} $ eV samples}
         \label{fig:hist_sb_1895}
     \end{subfigure}
     \hfill
     \begin{subfigure}[b]{0.31\textwidth}
         \centering
         \includegraphics[width=\textwidth]{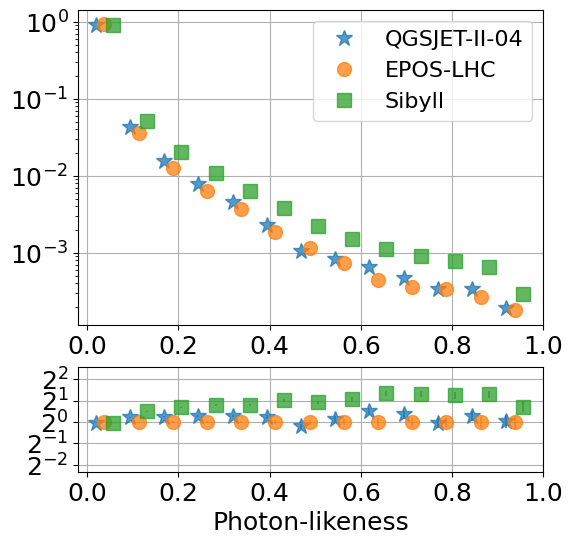}
         \caption{MC $ 10^{19.45} $ eV samples}
         \label{fig:hist_sb_1945}
     \end{subfigure}

    \text{\small{Without Sibyll-simulated events in the training phase}}
     \bigskip

    \begin{subfigure}[b]{0.31\textwidth}
         \centering
         \includegraphics[width=\textwidth]{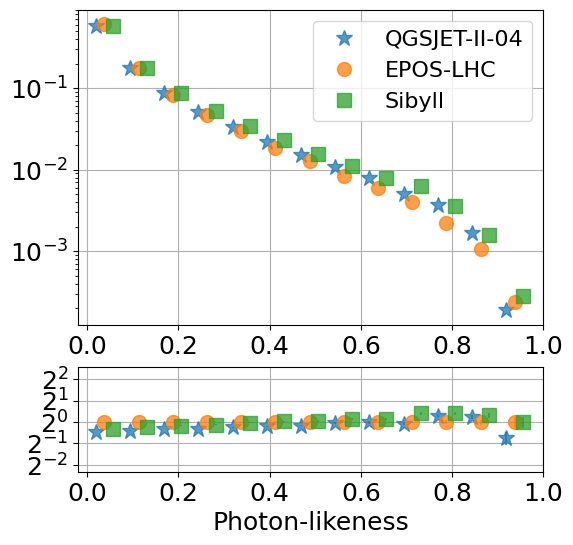}
         \caption{MC $ 10^{17.45} $ eV samples}
         \label{fig:hist_all_1745}
     \end{subfigure}
     \hfill
     \begin{subfigure}[b]{0.31\textwidth}
         \centering
         \includegraphics[width=\textwidth]{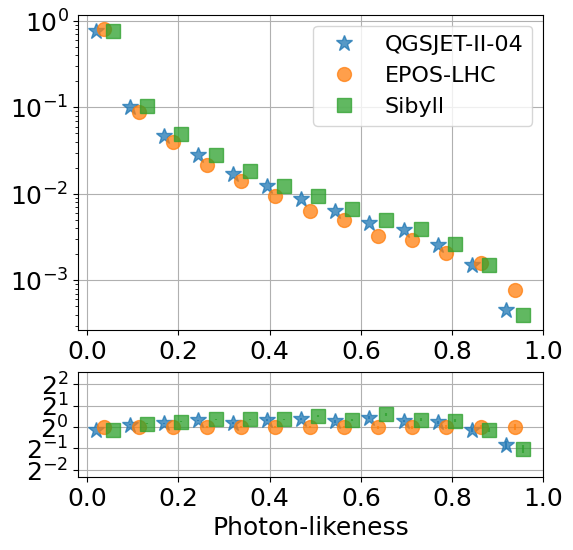}
         \caption{MC $ 10^{18.95} $ eV samples}
         \label{fig:hist_all_1895}
     \end{subfigure}
     \hfill
     \begin{subfigure}[b]{0.31\textwidth}
         \centering
         \includegraphics[width=\textwidth]{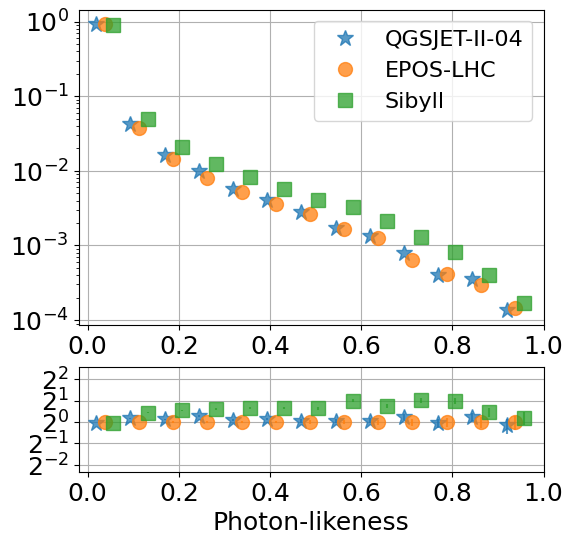}
         \caption{MC $ 10^{19.45} $ eV samples}
         \label{fig:hist_all_1945}
     \end{subfigure}
     \text{\small{With Sibyll-simulated events in the training phase}}
     
    \caption{Comparison of the histograms of neural network predictions on proton-induced MC samples simulated using different high-energy hadronic interaction models. The vertical axis shows the number of events in the corresponding bin. The lower subplots depict the ratios of events in the histogram bins with respect to events simulated using EPOS-LHC. To avoid overlapping, points for different hadronic interaction models are slightly shifted horizontally.}
    \label{fig:hists_train_model_dependence}
\end{figure}

Figure \ref{fig:hists_train_model_dependence} presents histograms of neural network predictions for proton-induced events in the common the test data set. We use two samples, starting from $ 10^{17.45} $ eV and $ 10^{18.95} $ eV, to separately study high-energy events. For the MC $ 10^{17.45} $ eV and $ 10^{18.95} $ eV samples, the predictions of neural networks are similar for all hadronic interaction models for both neural networks. The sole notable exception is that EPOS-LHC produces more photon-like events in the rightmost bin for the $ 10^{18.95} $ eV sample. For the MC $ 10^{19.45} $ eV sample, the events simulated with Sibyll are more photon-like, up to a factor of two in each bin. This discrepancy is less pronounced, yet significant, for the neural network trained on all hadronic interaction models.

\begin{figure}
\center{\includegraphics[width=0.5\linewidth]{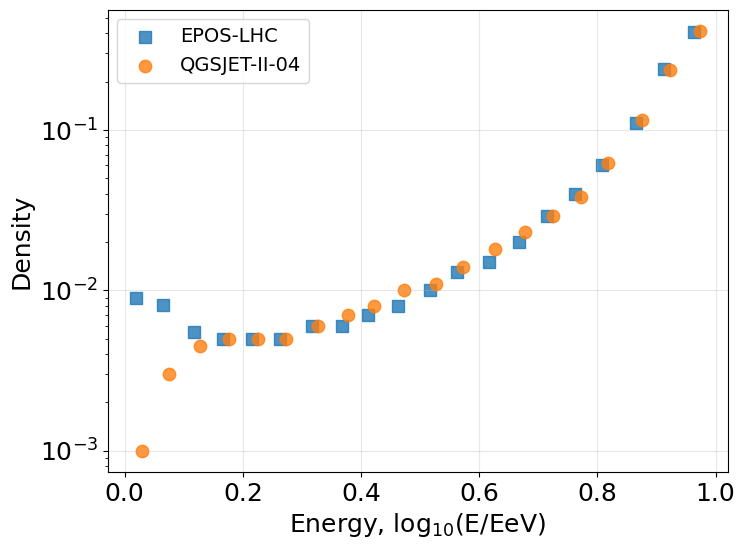}}\\
\caption{Comparison of neural network predictions for photon-induced air showers simulated using QGSJET-II-04 and EPOS-LHC hadronic interaction models.}
\label{gamma_models_compare}
\end{figure}

We also examined the model dependence of neural network predictions for photon-induced air showers.
Figure \ref{gamma_models_compare} illustrates the histograms of the neural network predictions, trained on QGSJET-II-04 MC only, for the events simulated using QGSJET-II-04 and EPOS-LHC.
The histograms exhibit a discrepancy for proton-like events ($ \xi<0.1 $) but are in good agreement for $ \xi>0.5 $.
This observation shows that the evolution of photon-induced air showers exhibits some dependence on the choice of hadronic interaction model.
However, for non-proton-like events ($ \xi>0.5 $), which are of primary interest for this study, the difference is negligible. We also tested that in the high photon-likeness region, $\xi>0.9$, photon exposure difference is below 7\% for all energy ranges.
This justifies our decision to use only one hadronic interaction model for simulating photon-induced air showers.

Based on these observations, we conclude that at energies below $ 10^{19.5} $ eV the effect of choosing a particular hadronic interaction model is negligible. At higher energies the effect becomes more pronounced.

Given that no single hadronic interaction model is inherently preferable, we opted to train neural networks using all available hadronic interaction models. This approach serves to mitigate the model dependence of the predictions and thus provides a robust upper limit on photon flux.

\section{Results}
\label{sec:results}

Now we present the results of applying the analysis described above to the TA SD data collected during 14 years of observation. 

\begin{figure}
 \begin{subfigure}{0.45\textwidth}
     \includegraphics[width=\textwidth]{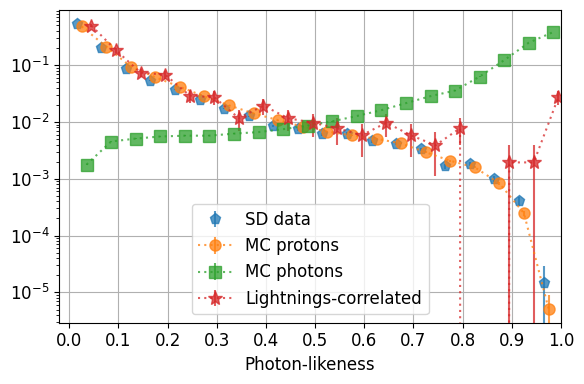}
     \caption{Energy $\geq ~ 10^{18.5}$ eV.}
     \label{fig:hist_1850}
 \end{subfigure}
 \hfill
 \begin{subfigure}{0.45\textwidth}
     \includegraphics[width=\textwidth]{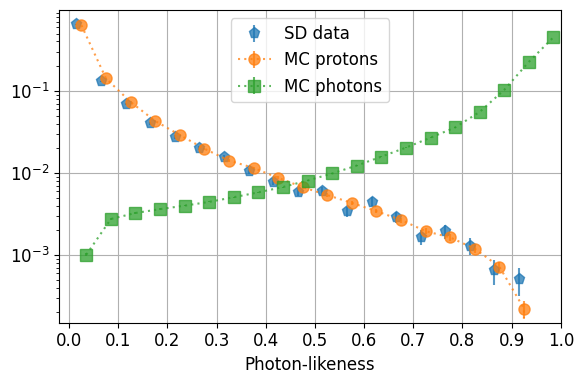}
     \caption{Energy $\geq ~ 10^{19.0}$ eV.}
     \label{fig:hist_1900}
 \end{subfigure}
 
 \medskip
 \begin{subfigure}{0.45\textwidth}
     \includegraphics[width=\textwidth]{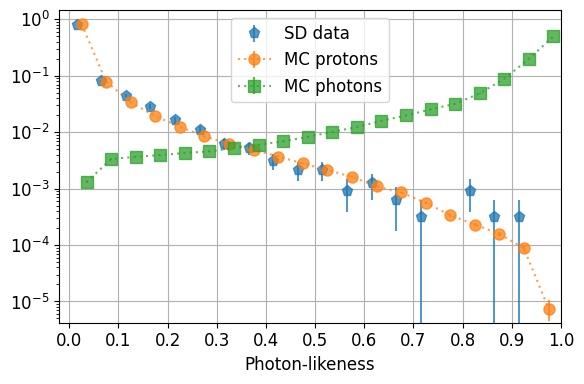}
     \caption{Energy $\geq ~ 10^{19.5}$ eV.}
     \label{fig:hist_1950}
 \end{subfigure}
 \hfill
 \begin{subfigure}{0.45\textwidth}
     \includegraphics[width=\textwidth]{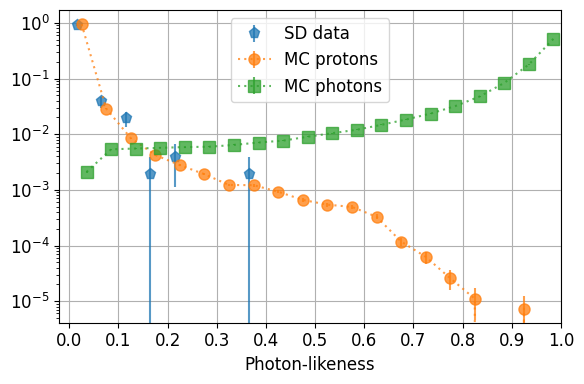}
     \caption{Energy $\geq ~ 10^{20.0}$ eV.}
     \label{fig:hist_2000}
 \end{subfigure}
 \caption{Histograms of neural network predictions for SD and MC data at various energy cuts. Error bars denote statistical uncertainties.}
 \label{fig:sd_hists}
\end{figure}

Figure \ref{fig:sd_hists} depicts histograms of neural network predictions for MC and experimental data across various energy thresholds. The predictions for proton MC and experimental data demonstrate good agreement. Importantly, there is no evidence of overfitting to the SD data in the region $ \xi \in [0, 0.2] $, which was used to fine-tune the neural network. 

The same distributions, obtained without fine-tuning the neural networks on experimental data, are shown in Figure \ref{fig:sd_hists_no_post} in Appendix \ref{app:no_fine}. The fine-tuning procedure significantly improved the agreement between proton MC and experimental data. This improvement is quantified by the mean asymmetric ratio between the histogram bins, which was reduced from 5.1 to 1.2 for events with $E>10^{18.5}$ eV.

As an additional validation measure, we examined the neural network's predictions on events correlated with lightning strikes registered near TA SD, Figure \ref{fig:hist_1850}. The corresponding histogram exhibits a peak near $ \xi =1 $, indicating that some of the events are indeed lightning-induced. We have also identified Terrestrial Gamma-Ray Flashes and lightning-correlated events previously reported by TA as photon-like with high confidence \cite{kieu2024first, Abbasi:2024vlj, TelescopeArrayProject:2017elp}. This confirms that the neural network correctly learned to identify photon-like events since lightning-induced events are essentially electromagnetic cascades, which are similar to photon-induced air showers \cite{kieu2024first}.

\begin{figure}
\center{\includegraphics[width=0.99\linewidth]{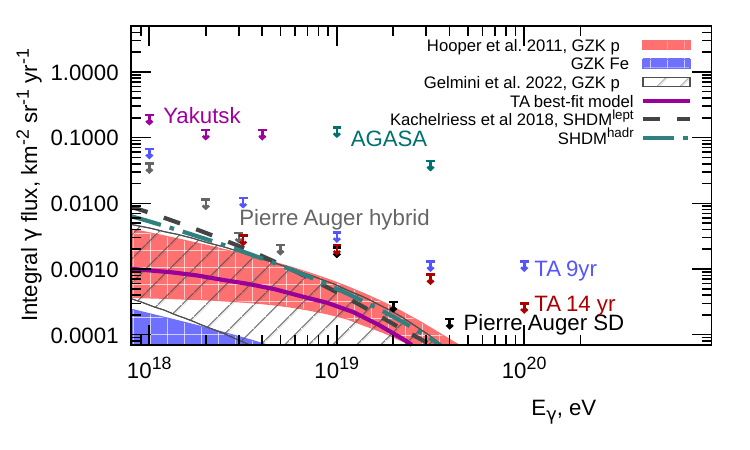}}\\
\caption{Upper limits on diffuse flux of UHE photons obtained in this study (TA 14 yr) compared with the previous TA result~\cite{TelescopeArray:2018rbt} and limits obtained by AGASA~\cite{Shinozaki:2002ve}, Yakutsk~\cite{Glushkov:2009tn}, and Pierre Auger hybrid~\cite{PierreAuger:2024ayl} and SD~\cite{abreu2023search}.
For comparison we show the expected flux of cosmogenic UHE photons for several UHECR flux models: pure protons (red band~\cite{Hooper:2010ze} and hatched band~\cite{Gelmini:2022evy}), pure iron (blue band)~\cite{Hooper:2010ze}, in both cases the bands denote the theoretical uncertainties of the models, and TA best-fit model~\cite{Bergman:2021djm, Kuznetsov:2025ehl}. The predictions for SHDM decaying via hadronic and leptonic channels are also shown~\cite{Kachelriess:2018rty}.
}
\label{img_results}
\end{figure} 

\begin{table}
\begin{tabularx}{\textwidth}{|W{1.}|W{1.}|W{1.}|W{1.}|W{1.}|}
 \hline
Energy,
log$_{10}$(E/1eV) & Quality cuts & Photon selection
efficiency, $S(\xi)$ & $\xi$-cut value & Effective exposure, $10^3 ~\text{km}^2 
\cdot \text{sr} \cdot \text{yr}$\\
\hline
18.5 & 22.8\% & 25.4\% & 0.95 & 0.98 \\
\hline
19.0 & 46.1\% & 45.9\% & 0.932 & 3.58 \\
\hline
19.5 & 58.2\% & 64.2\% & 0.909 & 6.30 \\
\hline
20.0 & 67.7\% & 91.4\% & 0.691 & 10.44 \\
\hline
\end{tabularx}
\caption{Contributions of the cuts to the effective exposure at each energy threshold. The value represents the ratio of the exposure after the given cut to the exposure before cut. The value of the optimal $\xi$-cut is given for reference.}
\label{table_exposure}
\end{table}

\begin{table}
\begin{tabularx}{\textwidth}{|W{1.}|W{1.}|W{1.}|W{1.}|W{1.}|W{1.}|}
 \hline
Energy, log$_{10}$(E/eV) ~ & Proton suppression level & Expected proton bg. & Observed photon candidates  & $ \Phi_\gamma $, $(\text{km}^2 \cdot \text{sr} \cdot \text{yr})^{-1}$ & Photon fraction upper limit\\
 \hline
 18.5 & 8.2$\cdot 10^{-6}$ & 0.35 & 0 & 3.2$\cdot 10^{-3}$ & 1.3$\cdot 10^{-3}$ \\ 
 \hline
 19.0 & 4.4$\cdot 10^{-5}$ & 0.62 & 3 & 2.3$\cdot 10^{-3}$ & 6.8$\cdot 10^{-3}$  \\ 
 \hline
 19.5 & 5.6$\cdot 10^{-5}$ & 0.18 & 1 & 8.2$\cdot 10^{-4}$ & 2.1$\cdot 10^{-2}$  \\ 
 \hline
 20.0 & 1.2$\cdot 10^{-4}$ & 0.06 & 0 & 3.0$\cdot 10^{-4}$ & 0.21 \\ 
 \hline
\end{tabularx}
\caption{Resulting metrics and limits on the photon flux $ \Phi_\gamma $. Effective exposure is evaluated after imposing both the event quality cuts and the $\xi$ cut. Proton suppression level is calculated by directly counting the number of proton-induced events below and above the $\xi$ cut.}
\label{table_results}
\end{table}

To derive upper limits on the photon flux $ \Phi_\gamma $, we use Eq.~\ref{upper-limit}. We calculate the effective TA SD exposure taking into account TA SD geometry, event selection and trigger efficiency, exclusion of events below the classification threshold, and the use of burn sample for fine-tuning the neural network. The efficiency of quality and $\xi$ cuts along with the effective exposure is shown in Table~\ref{table_exposure}.

The resulting limits on photon flux $ \Phi_\gamma $ are presented in Table~\ref{table_results}. We observed no significant excess of photon candidates compared to the expected proton background. The upper limits on the photon fraction were calculated as a ratio of the corresponding photon flux limit to the integral flux of the TA SD spectrum~\cite{TelescopeArray:2023bdy}. The energy, zenith angle and Linsley front curvature of the observed photon candidates are reported in Appendix \ref{app:cand_params}.

In Fig.~\ref{img_results} we compare the present limits with the limits of AGASA~\cite{Shinozaki:2002ve}, Yakutsk~\cite{Glushkov:2009tn} and Pierre Auger hybrid ~\cite{PierreAuger:2024ayl} and SD observations~\cite{abreu2023search}.
For comparison we depict the previous TA limits obtained using boosted decision trees for 9 years of SD operation time~\cite{TelescopeArray:2018rbt}.
In the same figure we also show theoretical predictions for the UHE photon flux. The predictions for cosmogenic photons (GZK photons) depend on the mass composition of initial UHECR flux: pure protons~\cite{Hooper:2010ze, Gelmini:2022evy} and pure iron nuclei~\cite{Hooper:2010ze} are shown as extreme cases. The uncertainties of the predicted photon fluxes due to uncertainties of the initial UHECR flux and its propagation are shown as bands. We also show the prediction of the cosmogenic UHE photon flux~\cite{Kuznetsov:2025ehl} for the UHECR flux model derived as a best fit for the TA spectrum and $X_{\rm max}$ data~\cite{Bergman:2021djm}. The expectations for UHE photon fluxes from the decay of super heavy dark matter (SHDM) particles $X$ are shown for two benchmark models: $X \rightarrow q\bar{q}$ and $X \rightarrow \nu\bar{\nu}$~\cite{Kachelriess:2018rty}. In both cases the $X$ mass, $M_X$, is $10^{12}$~eV and the its lifetime, $\tau_X$, is normalized to the Pierre Auger hybrid limits~\cite{PierreAuger:2024ayl}, which are the most stringent constraints for these scenarios. Specifically, we use $\tau_X = 7.1 \times 10^{22}$~yr for the $X \rightarrow q\bar{q}$ model and $\tau_X = 1.7 \times 10^{22}$~yr for the $X \rightarrow \nu\bar{\nu}$ model.

\section{Conclusions}
\label{sec:concl}

In this work, we have presented an updated search for the diffuse flux of ultra-high-energy photons using a 14-year data set from the Telescope Array Surface Detector. 
By employing a neural network classifier that combines reconstructed composition-sensitive observables with time-resolved SD waveforms, and fine-tuning it with a dedicated burn sample of experimental data, we achieved a robust and efficient separation between photon-induced and proton-induced events. Together with the increased statistics of the TA SD data set, this leads to a substantial improvement over our previous results, in particular at the energy thresholds of $10^{18.5}$ and $10^{20.0},\mathrm{eV}$, where the constraints are strengthened by approximately a factor of three. At intermediate energies, the sensitivity improvement is moderated by the observation of a small number of photon candidates, which remain consistent with statistical fluctuations of the expected hadronic background.
As a result, we derived stringent 95\% confidence level upper limits on the photon flux, which are the most strict in the Northern Hemisphere and complement the limits set by the Pierre Auger Observatory in the Southern Hemisphere. 

We found no significant excess of photon candidates above the expected hadronic background. Our limits approach the upper edge of the GZK photon predictions~\cite{Hooper:2010ze, Gelmini:2022evy}; however, probing the GZK photon models~\cite{Kuznetsov:2025ehl} will require an improvement of several times in sensitivity. Notably, in the search for signals from SHDM decay, our limits are competitive with those of Auger, despite the fact that TA does not observe observe the Galactic center region, which is abundant for the dark matter.
Looking ahead, we hope to support or constrain such models in future with increased TA SD exposure time or improved machine learning-based event selection criteria.

\section*{Code Availability}
Code for reproducing the architecture of the neural network described in section \ref{sec:nn} is available at \small{\url{https://github.com/ml-inr/TA-gamma-search}}.

\acknowledgments

The Telescope Array experiment is supported by the Japan Society for
the Promotion of Science(JSPS) through
Grants-in-Aid
for Priority Area
431,
for Specially Promoted Research
JP21000002,
for Scientific  Research (S)
JP19104006,
for Specially Promoted Research
JP15H05693,
for Scientific  Research (S)
JP19H05607,
for Scientific  Research (S)
JP15H05741,
for Science Research (A)
JP18H03705,
for Young Scientists (A)
JPH26707011,
for Transformative Research Areas (A)
JP25H01294,
for International Collaborative Research
24KK0064,
and for Fostering Joint International Research (B)
JP19KK0074,
by the joint research program of the Institute for Cosmic Ray Research (ICRR), The University of Tokyo;
by the Pioneering Program of RIKEN for the Evolution of Matter in the Universe (r-EMU);
by the U.S. National Science Foundation awards
PHY-1806797, PHY-2012934, PHY-2112904, PHY-2209583, PHY-2209584, and PHY-2310163, as well as AGS-1613260, AGS-1844306, and AGS-2112709;
by the National Research Foundation of Korea
(2017K1A4A3015188, 2020R1A2C1008230, and RS-2025-00556637) ;
by the Ministry of Science and Higher Education of the Russian Federation under the contract 075-15-2024-541, IISN project No. 4.4501.18, by the Belgian Science Policy under IUAP VII/37 (ULB), by National Science Centre in Poland grant 2020/37/B/ST9/01821, by the European Union and Czech Ministry of Education, Youth and Sports through the FORTE project No. CZ.02.01.01/00/22\_008/0004632, and by the Simons Foundation (MP-SCMPS-00001470, NG). This work was partially supported by the grants of the joint research program of the Institute for Space-Earth Environmental Research, Nagoya University and Inter-University Research Program of the Institute for Cosmic Ray Research of University of Tokyo. The foundations of Dr. Ezekiel R. and Edna Wattis Dumke, Willard L. Eccles, and George S. and Dolores Dor\'e Eccles all helped with generous donations. The State of Utah supported the project through its Economic Development Board, and the University of Utah through the Office of the Vice President for Research. The experimental site became available through the cooperation of the Utah School and Institutional Trust Lands Administration (SITLA), U.S. Bureau of Land Management (BLM), and the U.S. Air Force. We appreciate the assistance of the State of Utah and Fillmore offices of the BLM in crafting the Plan of Development for the site.  We thank Patrick A.~Shea who assisted the collaboration with much valuable advice and provided support for the collaboration's efforts. The people and the officials of Millard County, Utah have been a source of steadfast and warm support for our work which we greatly appreciate. We are indebted to the Millard County Road Department for their efforts to maintain and clear the roads which get us to our sites. We gratefully acknowledge the contribution from the technical staffs of our home institutions. An allocation of computing resources from the Center for High Performance Computing at the University of Utah as well as the Academia Sinica Grid Computing Center (ASGC) is gratefully acknowledged.
The lightning data used in this paper was obtained from Vaisala, Inc. We appreciate Vaisala’s academic research policy.

\appendix

\section{Reconstruction procedure}
\label{app:reco}
The reconstruction procedure employs a joint fit of the event geometry and lateral distribution function~\cite{TelescopeArray:2013yze}. The fit incorporates seven free parameters:
\begin{itemize}[noitemsep,topsep=0pt]
    \item[1-2)] Position of the shower core, $ \vec{r}_{\text{core}} = (x_{\text{core}},~y_{\text{core}},~0) $;
    \item[3-4)] Arrival direction of the primary particle: zenith ($ \theta $) and azimuth ($ \phi $) angles;
    \item [5)] Normalization of the shower's lateral distribution profile, $ S_{800} \, $;
    \item[6)] Time offset, $ t_0 \,$;
    \item[7)] Linsley front curvature parameter, $ a $. 
\end{itemize}
The fit optimizes two target functions: the SD stations activation times $t (\vec{r}) $ and their deposited charges $ S(r) $,
\begin{subequations} 
\begin{gather}
 t(\vec{r}) = t_0 + t_{\text{plane}}(\vec{r}) + a\times LDF(r)^{-0.5}\times\left(1+\frac{r}{R_l}\right)^{1.5} \;, \\
 S(r) = S_{800}\times LDF(r) \;,
\end{gather}
\end{subequations}
where $ r $ is the distance from the station to the shower core (i.e. the shortest distance between a point and a line), $ t_{\text{plane}}(\vec{r}) $ is the arrival time of the shower's plane front to the station located at $ \vec{r} $, and $ LDF(r) $ is the empirical lateral distribution profile introduced in the AGASA experiment \cite{teshima1986properties} and fine-tuned to better fit TA data~\cite{abu2013cosmic}:
\begin{subequations} 
\begin{gather}
 t_{plane}(\vec{r}) = \frac{1}{c}\times (\vec{n},(\vec{r}-\vec{r}_{\text{core}}))  \;, \\
 LDF(r) = f(r)/f(r_0) \,, ~~ r_0 = 800 ~\text{m} \;, \\
 f(r) = \left(\frac{r}{R_m}\right)^{-1.2} \left(1+\frac{r}{R_m}\right)^{-(\eta-1.2)} \left(1+\frac{r^2}{R_1^2}\right)^{-0.6} \;.
\end{gather}
\end{subequations}
Here $ c $ is the speed of light, $ \vec{n}=\vec{n}(\theta,\phi) $ is the unit vector along the direction of the shower axis, $ (\cdot,\cdot) $ stands for the scalar product, and
\begin{subequations} 
\begin{gather}
R_m = 90~\text{m} \,, ~~~ R_1 = 1000~\text{m} \,, ~~~ R_l=30~\text{m} \;, \\
\eta = 3.97 - 1.79(\text{sec}(\theta)-1) \;.
\end{gather}
\end{subequations}

The energy of the primary particle is estimated as a function of $S_{800},~\theta$, and $\phi$ using a lookup table obtained from the photon MC simulations~\cite{TelescopeArray:2018rbt}.

Below is the list of 14 reconstructed parameters~\cite{TelescopeArray:2018rbt} used as input data for the neural network: 

\begin{itemize}[noitemsep,topsep=0pt]
    \item[1)] Reconstructed Linsley front curvature parameter ($a$); 
    \item[2-3)] Zenith and azimuth angles  ($ \theta, ~\phi $);
    \item[4)] Signal density at 800 m from the shower core ($ S_{800} $);
    \item[5)] $ \chi^2/d.o.f. $ for the joint geometry and LDF fit;
    \item[6-7)] Area-over-peak of the signal at 1200 m and the corresponding slope parameter \cite{abraham2008upper};
    \item[8-9)] $ S_b \equiv \sum_i \left( S_i\times\left(\frac{r_i}{r_0}\right)^b \right) $ for two values of $b$: 2.5 and 4.0. Here $ S_i $ and $ r_i $ are the integral signal of $i$-th station and its distance to the shower core, respectively, and $ r_0 = 1200~m $~\cite{Ros:2011zg}, corresponding to TA SD stations spacing;
    \item[10)] Sum of integral signals from all of the triggered stations;
    \item[11)] Asymmetry of the signal between the upper and lower layers of stations \cite{abbasi2019mass};
    \item[12)] Total number of peaks within all stations \cite{abbasi2019mass};
    \item[13)] Number of peaks for the station with the largest signal;
    \item[14)] Minimal distance between the reconstructed air-shower core and SD array edge.
\end{itemize}

\section{Results without fine-tuning}
\label{app:no_fine}

\begin{figure}
 \begin{subfigure}{0.45\textwidth}
     \includegraphics[width=\textwidth]{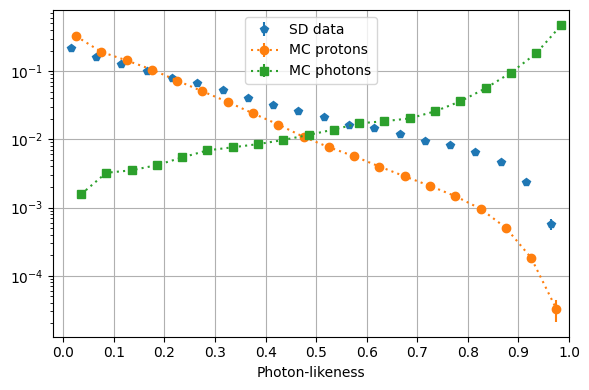}
     \caption{Energy $\geq ~ 10^{18.5}$ eV.}
     \label{fig:a}
 \end{subfigure}
 \hfill
 \begin{subfigure}{0.45\textwidth}
     \includegraphics[width=\textwidth]{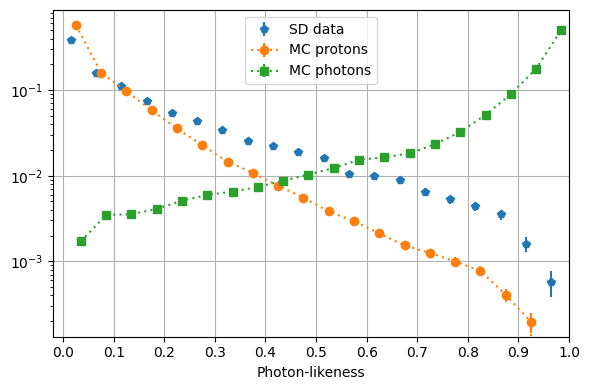}
     \caption{Energy $\geq ~ 10^{19.0}$ eV.}
     \label{fig:b}
 \end{subfigure}
 
 \medskip
 \begin{subfigure}{0.45\textwidth}
     \includegraphics[width=\textwidth]{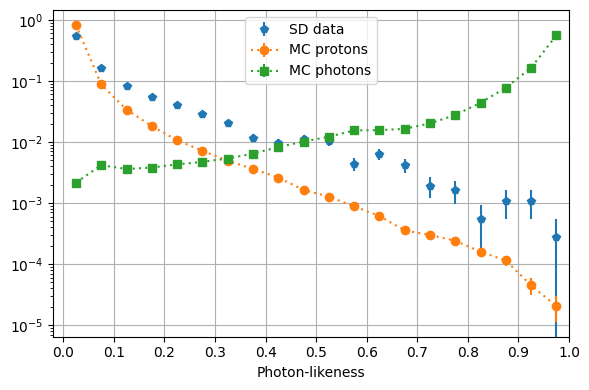}
     \caption{Energy $\geq ~ 10^{19.5}$ eV.}
     \label{fig:c}
 \end{subfigure}
 \hfill
 \begin{subfigure}{0.45\textwidth}
     \includegraphics[width=\textwidth]{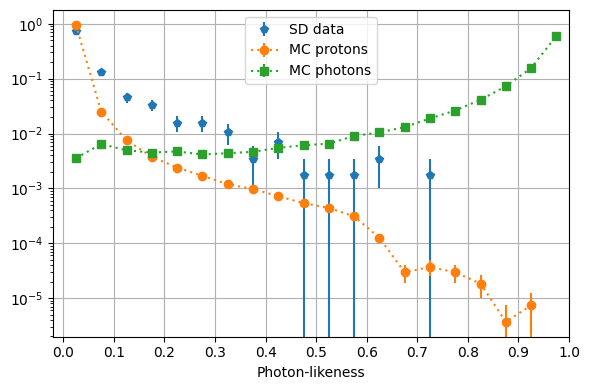}
     \caption{Energy $\geq ~ 10^{20.0}$ eV.}
     \label{fig:d}
 \end{subfigure}
 \caption{Histograms of neural network prediction for SD and MC data at various energy cuts without fine-tuning using experimental data. Error bars denote statistical uncertainty.}
 \label{fig:sd_hists_no_post}
\end{figure}

Figure \ref{fig:sd_hists_no_post} presents histograms of neural network predictions on experimental and MC simulated data without the fine-tuning process described in the main text. In this case, SD events exhibit more photon-like characteristics than MC data. This discrepancy can be due to small differences between MC and experimental data that are irrelevant for the standard reconstruction, but are important for the neural network. Fine-tuning the neural network on SD data mitigates these differences, resulting in more reliable predictions.

Our analysis revealed that the weighting scheme applied to proton- and photon-induced events during neural network training influences the consistency of MC and experimental data histograms. Specifically, assigning higher weights to photon events tends to increase the similarity between these histograms. However, we retain our original weighting ratio for two reasons. Firstly, this weighting strategy compels the neural network to minimize false-positive photon identifications, which is crucial for setting stronger limits on photon flux. Secondly and more importantly, altering the weights based on observed outcomes would compromise the blind nature of our analysis.

\section{Photon candidates}
\label{app:cand_params}

Below we present the main reconstructed parameters of the observed photon candidates. Note that their photon-likeness is very close to the $\xi$-cut value for $10^{19.5}$ eV limit.

\begin{table}
\begin{tabularx}{\textwidth}{|W{1.65}|W{1.8}|W{0.9}|W{0.75}|W{0.5}|W{0.8}|W{0.8}|W{0.8}|}
 \hline
 Date (yyyy.mm.dd)  & Time \newline (hhmmss.usec) & $\xi$-value
 & $E_\gamma $, EeV & $E_p$, EeV & $\theta_\gamma$ & $\phi_\gamma$ & $a_{curv}$ \\
\hline
2011.08.02 & 055738.387354 & 0.934 & 11.57 & 5.08 & 23.0\degree & 208.7\degree & 1.44 \\
\hline
2011.10.02 & 235948.253047 & 0.939 & 37.29 & 13.57 & 15.1\degree & 110.9\degree & 0.75 \\
\hline
2015.07.09 & 052937.805091 & 0.935 & 11.06 & 4.61 & 36.0\degree & 11.6\degree  & 0.98 \\
\hline
\end{tabularx}
\caption{Parameters of the observed photon candidates. $E_\gamma$ and $E_p$ are event energies reconstructed using photon-optimized and proton-optimized algorithms, $a_{curv}$ stand for Linsley front curvature. The reconstructed zenith and azimuth angles are given with respect to the standard TA SD coordinate system.}
\end{table}

\bibliography{ref_ta_gamma}

\end{document}